\documentclass[twocolumn]{aastex62}

\usepackage{color}

\shorttitle{The formation of Milky Way globular cluster system}
\shortauthors{A.E. Piatti}

\begin{document}

\title{Formation imprints in the kinematics of the Milky Way globular cluster system}

\author[0000-0002-8679-0589]{Andr\'es E. Piatti}
\affiliation{Consejo Nacional de Investigaciones Cient\'{\i}ficas y T\'ecnicas, Godoy Cruz 2290, C1425FQB, 
Buenos Aires, Argentina}
\affiliation{Observatorio Astron\'omico de C\'ordoba, Laprida 854, 5000, 
C\'ordoba, Argentina}
\correspondingauthor{Andr\'es E. Piatti}
\email{e-mail: andres.piatti@unc.edu.ar}

\begin{abstract}
We report results on the kinematics of Milky Way (MW) globular clusters (GCs)
based on updated space velocities for nearly the entire
GC population.  We found that a 3D space with the semi-major axis, the eccentricity
and the inclination of the orbit with respect to the MW plane as its axes is helpful in 
order to dig into the formation of the GC system. We find that GCs formed {\it in-situ} 
show a clear correlation between their eccentricities and their orbital inclination
in the sense that clusters with large eccentricities also have large inclinations.
These GCs also show a correlation between their distance to the MW center
and their eccentricity. Accreted GCs  do not exhibit 
a relationship between eccentricity and inclination, but span a wide variety of inclinations at 
eccentricities larger than $\sim$ 0.5. Finally, we computed the velocity anisotropy $\beta$ 
of the GC system and found for GCs formed {\it in-situ}
that $\beta$ decreases from $\approx$ 0.8 down to 0.3 from the outermost regions towards the MW
center, but remains fairly constant (0.7-0.9) for accreted ones.
These findings can be explained if GCs formed from gas that collapsed radially 
in the outskirts, with preference for relative high infall angles. 
As the material reached the rotating forming disk, it became more circular and
moved with lower inclination relative to the disk. A half of the GC population was accreted and deposited in orbits covering the entire range of energies from the outer 
halo to the bulge.
\end{abstract}


\section{Introduction}

The study of the orbital motion of Milky Way (MW) globular clusters (GCs) has gained
a renewed enthusiasm since the second data release (DR2) of the {\it Gaia} mission 
\citep{gaiaetal2016,gaiaetal2018b} became publicly available 
\citep[see, e.g.][]{lietal2018,s2019,watkinsetal2019}. Previous studies of MW GC motions 
have shed some light on our knowledge about their formation and assembly history.
For instance, \citet{dinescuetal1999} obtained orbits for 38 GCs and 
found that some of them have large eccentricities and apogalactic distances larger
than 10 kpc. They also found that internal two-body relaxation is more important than 
the destruction processes due to disk and bulge shocking. More recently,
\citet{perezvillegasetal2018} analyzed 9 bulge GCs and concluded that they move on 
rather eccentric prograde or retrograde orbits that are strongly influenced by the Galactic bar. A
chaotization of the cluster orbits due to the MW bar was also found by \citet{chemeletal2018} from the analysis 
of the motions of 115 GCs in a non-axisymmetric MW potential with a bar.

As far as we are aware, the most complete compilation of {\it Gaia} DR2 proper motions
and ground-based line-of-sight velocities to date is that of 
\citet{baumgardtetal2019}\footnote{Available at: https://people.smp.uq.edu.au/HolgerBaumgardt/globular/}, 
who derived from them the space velocities of 156 GCs, and  velocity 
dispersion profiles of 141 GCs. Their data set includes all GCs analyzed
by \citet{gaiaetal2018a} and \citet{vasiliev2019}, respectively.
\citet{baumgardtetal2019} derived the total mass lost by GCs since 
their formation by computing their orbital motions backwards in time, accounting
for mass-loss and dynamical friction. They found that the dynamical evolution plays
an important role in the GC's mass loss process, in agreement with \citet{dinescuetal1999}.
The derived  Galactic positions ($X,Y,Z$), space velocities ($U,V,W$) and 
perigalactic ($R_{peri}$) and apogalactic ($R_{apo}$) distances can now be exploited
to go forward in our understanding  
of the dynamical behavior of the ancient Galactic
GC system,  and hence to draw some clues on the formation of the MW.

Precisely, in this work we comprehensively analyze the positions and velocities obtained
by \citet{baumgardtetal2019}, and discuss the relationship between different orbital properties,
in order to unveil possible scenarios of the events that took place during
the formation of the MW. In Section 2 we derive the aforementioned kinematic properties, whereas
Section 3 deals with the analysis of some relevant relationships and the comparison with
recent results on different mechanisms of the GC formation. Finally, we summarize the main 
conclusions of this work in Section~4.

\section{Orbital properties}

Several kinematic properties can be derived from the orbital parameters obtained by
\citet{baumgardtetal2019}, namely: from the average values of $R_{peri}$ and  $R_{apo}$ we
define the mean semi-major axis of the GCs'  orbits as:

\begin{equation}
a =   \frac{R_{peri} +  R_{apo}}{2}.
\end{equation}

The semi-major axis
has the advantage to be less time-dependant than the GC's Galactocentric distance 
($R_{GC}$), and is more representative of the distance of a GC's birthplace to
the Galactic center or the average distance where a GC was deposited after accretion
of its host dwarf galaxy onto the MW. We also computed the orbital eccentricity ($\epsilon$)
as:

\begin{equation}
\epsilon = \frac{R_{apo} - R_{peri}}{R_{apo} + R_{peri}};
\end{equation}

\noindent the components of the angular momentum:

\begin{equation}
L_X  =  Y\times W - Z\times V,
\end{equation}
\begin{equation}
L_Y = Z\times U - X\times W,
\end{equation}
\begin{equation}
L_Z = X\times V - Y\times U;
\end{equation}

\noindent and the inclination of the orbit:

\begin{equation}
i = acos\left(\frac{L_Z}{\sqrt{L_X^2 + L_Y^2 + L_Z^2}}\right).
\end{equation}

Note that $i$ values range  from 0$\degr$ for fully prograde in-plane orbits to 90$\degr$ for polar orbits
to 180$\degr$ for in-plane retrograde orbits.

We also transformed the $U$, $V$ and $W$ space velocity components to
the spherical ones $V_r$, $V_{\theta}$ and $V_{\phi}$.


 


For each orbital property $f(x_1,x_2,...,x_n)$, we derived its respective uncertainty
through Monte Carlo simulations. We run one thousand computations of 
$f(x_1,x_2,...,x_n)$ for each GC, each time using random values for
all the involved independent variables $x_i$ ($i=1,...,n$). These random
values were choosing among all possible ones
in the interval [$<x_i> -  \sigma(x_i)$, $<x_i> +  \sigma(x_i)$], where $<x_i>$ and  $\sigma(x_i)$ are the mean values
 and errors of the involved cluster properties (variables) derived by \citet{baumgardtetal2019}. 
 Then, we built a histogram from
 all the resulting $f(x_1,x_2,...,x_n)$ values and considered as the uncertainty of
  $f(x_1,x_2,...,x_n)$ 1/2 of the $f$ range where more than 16$\%$ and less than
  84$\%$ of the points are distributed.



Finally, we calculated the velocity anisotropy $\beta$. 
In doing this, we have first split the GC sample into three groups:
$a$ $\le$ 3 kpc (bulge); 3 kpc $<$ $a$ $\le$ 20 kpc (disk) and $a$ $>$ 20 kpc (outer
halo). Then, we  computed the velocity dispersions $\sigma V_r$, $\sigma V_\theta$
and $\sigma V_\phi$ in $V_r$, $V_\theta$ and $V_\phi$, respectively, using
a maximum likelihood approach by optimising the probability $\mathcal{L}$ that 
the sample of selected GCs with velocities $V_i$ and errors $e_i$ are 
drawn from a population with mean $<V>$ and dispersion $\sigma$  
\citep[e.g.,][]{pm1993,walker2006}, as follows:

\begin{equation}
\mathcal{L}\,=\,\prod_{i=1}^N\,\left( \, 2\pi\,(e_i^2 + \sigma^2 \, ) 
\right)^{-\frac{1}{2}}\,\exp \left(-\frac{(V_i \,- <V>)^2}{2(e_i^2 + \sigma^2)}
\right),
\end{equation}

\noindent where the errors on the mean and dispersion were computed from the respective 
covariance matrices. Finally, we computed $\beta$ as follows:

\begin{equation}
\beta = 1 - \frac{(\sigma V_\theta)^2 + (\sigma V_\phi)^2}{2 (\sigma V_r)^2},
\end{equation}

\begin{figure}
\includegraphics[width=\columnwidth]{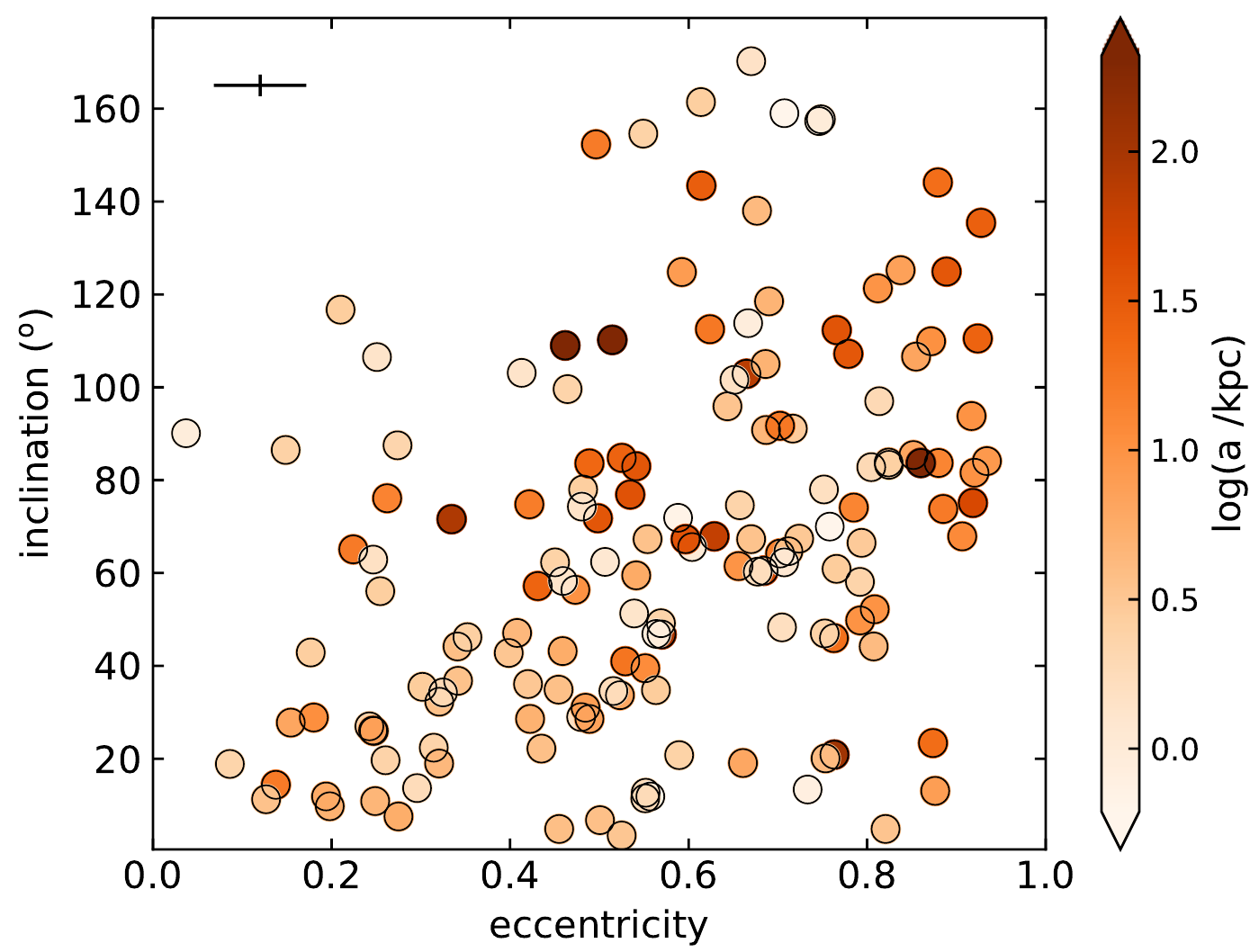}
\caption{Relationship between the inclination ($i$) and the eccentricity ($\epsilon$) for
the GC sample. Symbols have been
colored according to the color bar at the right margin. Typical error bars
are also indicated.}
 \label{fig1}
\end{figure}

We tried different relationships between the derived independent parameters, and found that
using  $i$ versus $\epsilon$ versus log($a$) results in the best enlightenment of 
the overall kinematic 
state of the GC system. Fig.~\ref{fig1} depicts this relationship for the  GC sample. As can be
seen, GCs with prograde orbits
do not span the whole ranges of $i$ and $\epsilon$ values randomly, but follow a general trend,
in such a way that $i$ increases with $\epsilon$. There are a handful of  GCs with
prograde orbits
with $\epsilon$  $\ga$ 0.5  and $i$ $\la$ 25$\degr$ that depart from this general relation, 
as well as some few GCs with $\epsilon$ $\la$ 0.3 and $i$ $\ga$ 50$\degr$.
Regardless of these cases, the unveiled  correlation shows that at a fixed eccentricity,
 the $i$ range can vary   (full range)  between $\Delta$($i$) $\sim$ 20$\degr$ 
 ($\epsilon$ $\sim$ 0.2)   and $\Delta$($i$) $\sim$ 80$\degr$ ($\epsilon$ $\sim$ 0.9).

We interpret this behavior as if the present-day inclinations -- along with the eccentricities and 
the semi-major axes -- of the GC population have somehow kept
imprints of their formation epoch.
In general, they have orbits with large $e$ and $i$ values.
For GCs with  prograde orbits and 3 kpc $<$ $a$ $\le$ 20 kpc, we derived a Spearman rank-order 
coefficient of 0.62 between their inclinations and their eccentricities, for those with $\epsilon$ $<$ 0.5 and $i$ $<$ 50$\degr$, 0.44, and for the whole 
prograde GC sample, we obtained 0.39. 
As for retrograde GCs, they have orbits with $\epsilon$ $\ga$ 0.5 and with
larger $i$ values as their semi-major axes increase. 

\begin{figure}
\includegraphics[width=\columnwidth]{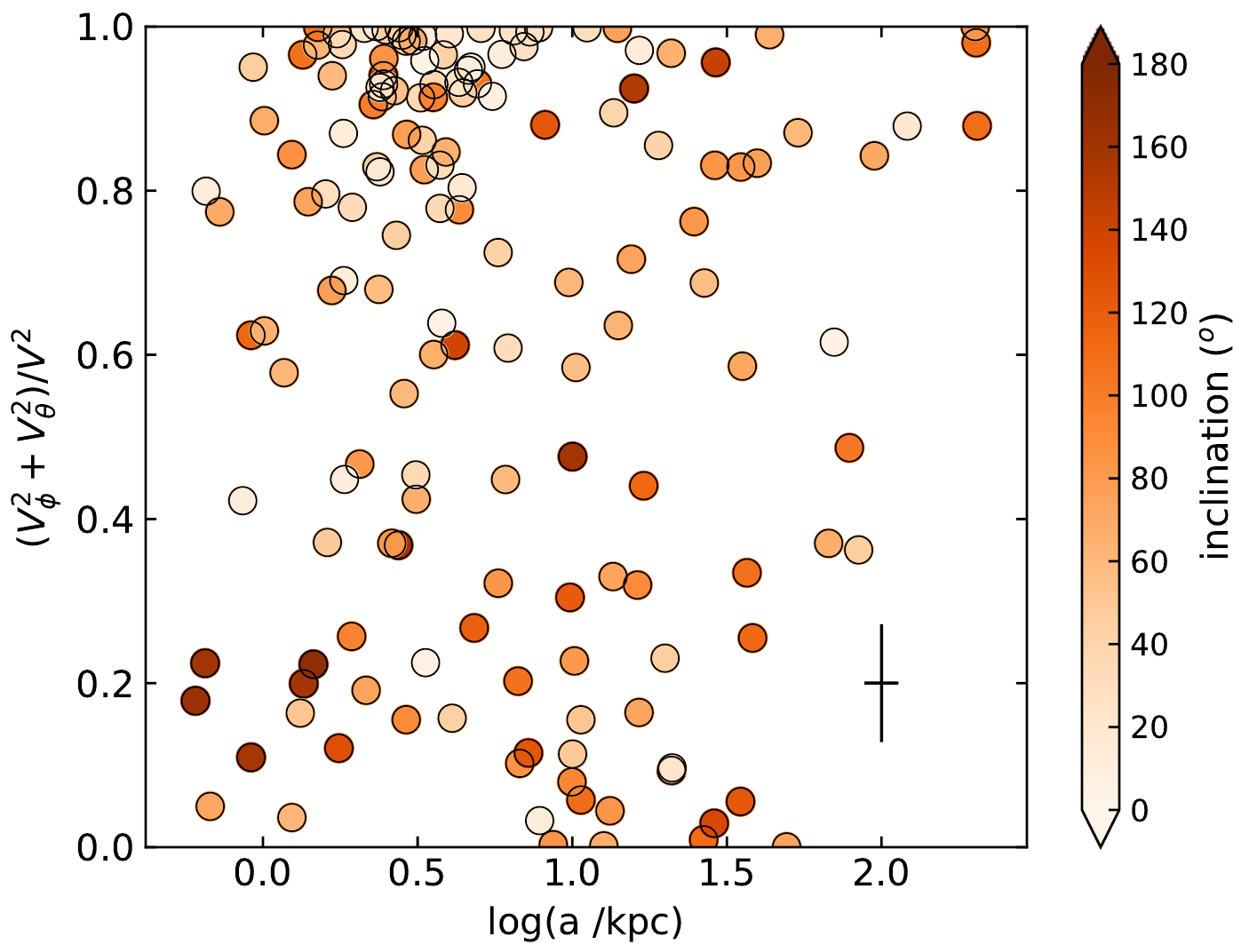}
\caption{ $(V_\phi^2 + V_\theta^2)/V^2$ ratio as a function of the
semi-major axis for the GC sample. Typical error bars are also drawn.}
 \label{fig2}
\end{figure}

These features reveal that, independently of the direction of rotation (prograde or retrograde), GCs with more circular orbits tend to be more numerous
in the inner parts than those in the outer parts of the Milky Way.
In order
to confirm such an orbital motion pattern, we plot in Fig.~\ref{fig2} the 
$(V_\phi^2 + V_\theta^2)/V^2$ ratio as a function of the
semi-major axis. It 
shows that  GCs with  log($a$ /kpc) $\la$ 0.8 kpc are more numerous for
$(V_\phi^2 + V_\theta^2)/V^2$ $\ga$ 0.8. Note that this behavior is observed in GCs rotating in prograde and retrograde orbits.

The transition from nearly radial orbits of the outermost GCs to more or less disk-like
rotating GCs in the MW main body (3 kpc $<$ $a$ $\le$ 20 kpc) to the
orbital  anisotropy of the MW bulge GCs is also supported by the variation of
the velocity anisotropy in terms of the distance  from the Galactic center. 
We  used the  computed $\sigma V_r$, $\sigma V_\theta$
and $\sigma V_\phi$ values as described above and then evaluated eq. (8). The resulting $\beta$ values
for the three distance ranges (bulge, disk, outer halo) turned out to
be 0.29, 0.51 and 0.79 for prograde orbits of GCs formed {\it in-situ} (see also Section 3 for a
discussion of GCs formed {\it in-situ}) and 0.72, 0.67 and 0.90 for retrograde
ones, respectively, with typical $\sigma(\beta)$ $\approx$  0.1. This result
shows that while prograde orbits of GCs formed {\it in-situ} lose the radial
imprints from the outer halo towards the bulge, the retrograde (accreted GCs,  see Section 3) ones keep it
throughout the whole MW.

\section{Analysis and discussion}

Nearly 75 per cent of the GCs have prograde orbits and they have mean $i$ values
of (70$\pm$20)$\degr$ and (30$\pm$20)$\degr$, for outer halo and disk GCs, respectively. 
On the other hand, most of the GCs with retrograde orbits -- thought to come from an accretion origin --  have been able to keep  their relatively high inclinations and
eccentricities, even though some of them have reached very deep into the central 
regions of the MW (see Figs.~\ref{fig1} and \ref{fig2}). 
Here we make  only use of the notion of accreted GCs described by \citet{fb2010} based on that
retrograde motions are the signature of objects that have been accreted in the opposite rotational sense to the 
main bulk of Milky Way’s rotation. Note that accreted GCs can also have prograde orbits, which we
considered in the following analysis. For this reason, \citet{fb2010} also investigated the age-metallicity 
relationship as a diagnostic tool
to disentangle accreted and formed {\it in-situ} GCs \citep[see also][]{kruijssenetal2018}.
Since accreted GCs
could also have prograde orbits, we assumed for them a similar number of objects as the GCs
with retrograde orbits and with their same inclination distribution. Thus, we were able to
subtract from the observed inclination distribution of prograde GCs that of
retrograde GCs to obtain the distribution of prograde GCs formed {\it in-situ}.

\begin{figure}
     \includegraphics[width=\columnwidth]{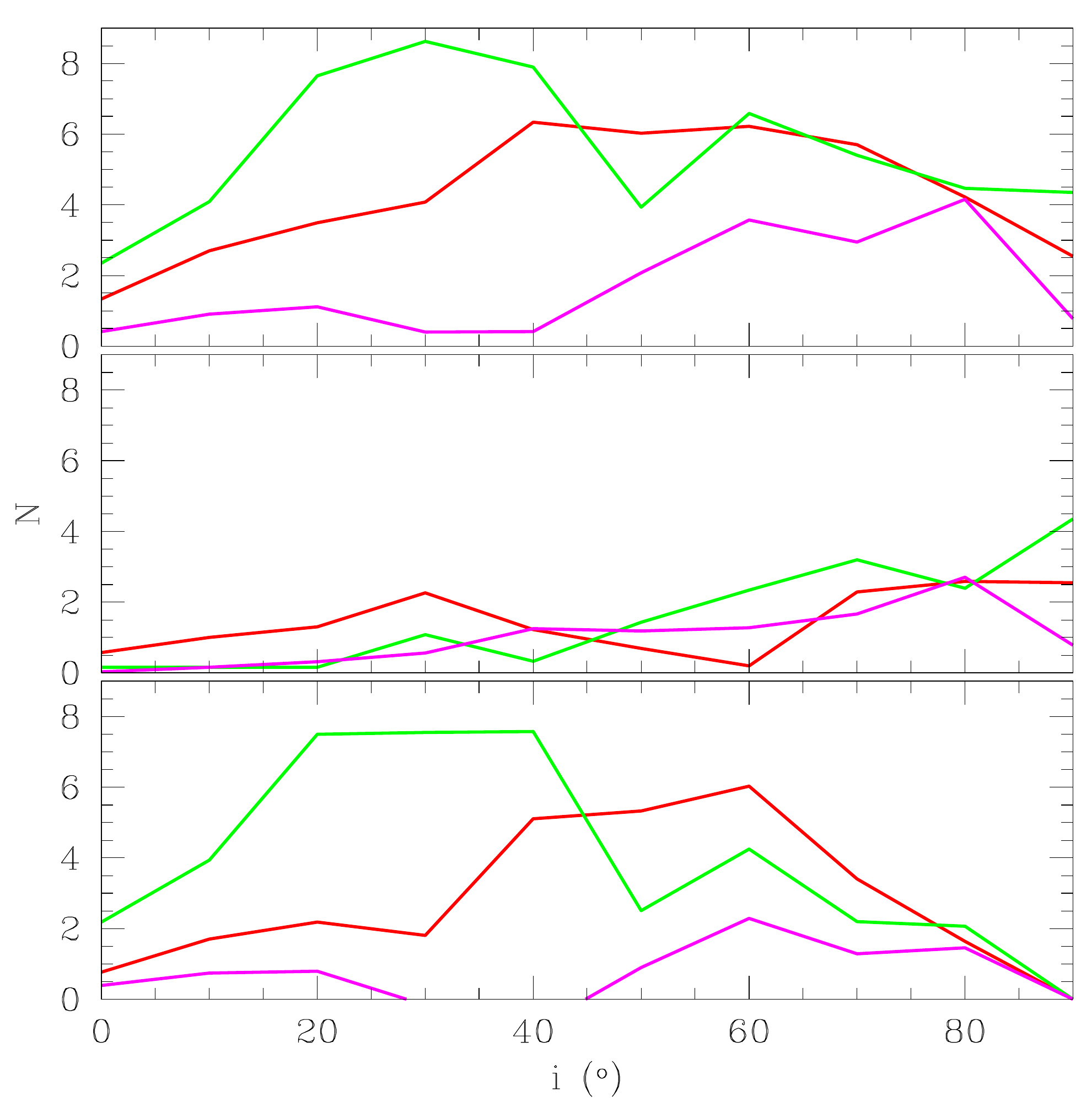}
\caption{Distribution of observed prograde (top panel) and  mirrowed (180$\degr$-$i$) retrograde (middle panel) GCs and prograde GCs formed {\it in-situ} (bottom panel)
with semi-major axes ($a$) in the ranges: $a$ $\le$ 3 kpc (red line), 3 kpc $<$ $a$ $\le$ 20 
kpc (lime line) and $a$ $>$ 20 kpc (magenta line), respectively.}
 \label{fig3}
\end{figure}

We computed these inclination distributions of GCs in prograde and retrograde orbits for 
our three semi-major axis ranges. In order to build those distributions we considered 
each inclination as represented by a one-dimensional {\it Gaussian} of unity area centered at the 
respective $i$ value, with a FWHM/2.355 equals to the $i$ error. 
Then, we used a grid of $i$ bins with sizes of $\Delta$($i$) = 10$\degr$  and added the fractions 
of the {\it Gaussian}s' areas that fall into the bin  boundaries. 
Thus, by taking into account the 
uncertainties of the $i$ estimates, we were able  to produce 
 actual observed $i$ distributions (not  distributions coming from considering only
mean individual $i$ values)
 \citep[see, e.g.][]{p14b,piattietal2018a}.
Fig.~\ref{fig3} shows the resulting distributions. As can be seen, 
there is a general different distribution of GCs in prograde orbits formed {\it in-situ} (bottom
panel) with respect to those on
retrograde GCs (middle panel). The latter show nearly similar inclination distributions in the three
spatial ranges.

 If we assumed that accreted GCs arrived uniformly from arbitrary directions, the distribution of their orbit poles would be uniform on a sphere. This means that the
number of points  at orbit poles $\sim$ 90$\degr$ ($i$ $\sim$ 0$\degr$) should be
 smaller than  that for orbit poles $\sim$ 0$\degr$ ($i$ $\sim$ 90$\degr$).
 Precisely, Fig.~\ref{fig3} (middle panel) shows -- at least for disk and outer
 halo GCs -- that the larger the inclination, the larger the number of GCs,
giving some support to the above assumption.

The outcome for GCs formed {\it in-situ} would imply that the initial collapse of gas out of which the 
GCs were formed 
was more or less isotropic -- outer halo GCs have orbits spanning the whole range
of inclinations --, and that, after the first disk passage, the motion of the gas
became more circular and  parallel to the Galactic plane -- disk GCs have orbits with
inclinations peaked at $\sim$ 30$\degr$. In the bulge region, opposite currents of 
gas clashed, resulting in GCs spanning the whole range of eccentricities and inclinations
(see also Figs.~\ref{fig1} and \ref{fig2}).  Therefore, there could be a transition from mostly radial (outer MW) 
to more circular (inner MW) prograde orbits. Note that the formation of all these prograde GCs
has happened in a space of time of $\la$ 3 Gyr \citep[][and references therein]{kruijssenetal2018}.
The accretion of GCs could have happened concurrently with the GC formation or a couple of Gyrs later \citep{helmietal2018}. Nevertheless, in either case, accreted GCs have not been fully subject
of the angular momentum acquired by the early MW disk.

 We have searched the literature seeking for any recent comprehensive model
of the MW GC formation and found that most of the latest developments
do not include kinematical GC signatures 
\citep[see, e.g.][]{renaudetal2017,elbadryetal2019}. \citet{bw2017} developed a 
model of the
hierarchical assembly of GCs and found that halo GCs show clearer rotation than
their stellar counterpart; \citet{fattahietal2019} showed that
halo metal-rich stars have highly eccentric orbits.  The outcomes of
\citet{bw2017}   agree with our finding of more eccentric orbits for
halo GCs. Nevertheless, the authors mentioned that their results are preliminary 
and that their analysis should be revisited.

Accretion of GCs has recently been more extensively discussed in the literature. 
\citet{helmietal2018} and \citet{belokurovetal2018} claimed that only one major merger with a dwarf 
galaxy slightly more massive than the Small Magellanid Cloud was responsible for the formation of 
the MW think disk $\sim$ 10 Gyr ago, while  \citet{pfefferetal2018} and \citet{kruijssenetal2018} 
introduced the accretion origin of GCs in a general cosmological context. Particularly, \citet{kruijssenetal2018} 
 found
that the MW has experienced no major mergers since $\sim$ 13 Gyr ago. Recently, \citet{gallartetal2019}
showed that there exist also an {\it in-situ} inner halo formed within the seed progenitor of the MW, 
just after the accreted inner halo population.

\begin{figure*}
     \includegraphics[width=\textwidth]{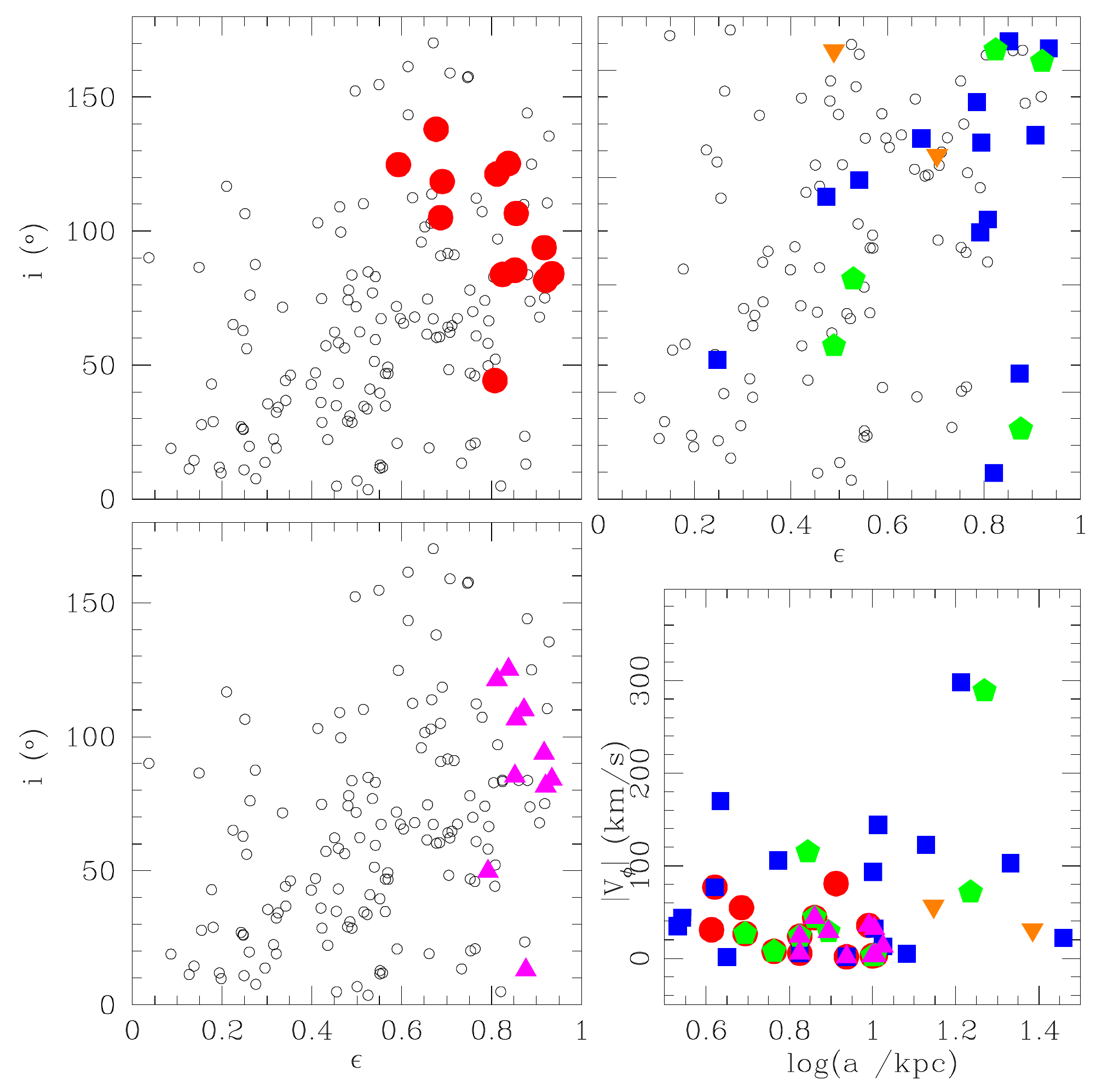}
\caption{Relationship between the inclination ($i$) and the eccentricity ($\epsilon$) for
 the GC sample.
GCs associated to Gaia-Enceladus (top-left panel, red symbols), Kraken, Canis Major and
Sagittarius (top-right panel. blue, lime and orange symbols, respectively) and Gaia-Sausage
(bottom-left panel, magenta symbols) are indicated. The bottom-right panel shows  only
colored symbols in the three other panels (see text in Sect. 3).}
 \label{fig4}
\end{figure*}

\citet{helmietal2018} associated to the merging dwarf galaxy Gaia-Enceladus 13 GCs 
(NGC\,288, 362, 1851, 1904, 2298, 4833, 5139, 5286, 6205, 6341, 6779, 7089, 7099)
with $L_z$ $<$ 250 kpc km/s, no mention whether their orbits are prograde or 
retrograde, but simply that they show a consistent age-metallicity relationship.
 However,  the top-left panel of Fig.~\ref{fig4} highlights the
positions of these GCs in the $i$ versus $\epsilon$ plane,  revealing that they have
relatively large eccentricities and  rotate in either prograde or retrograde orbits. 
In the case of
\citet{kruijssenetal2018}, the authors identified three less massive dwarf progenitors 
each with a number of GCs associated to them, namely: Sagittarius (NGC\,5634, 6715), Canis Major
(NGC\,1851, 1904, 2808, 4590, 5286, 6205, 6341, 6779, 7078, IC 4499) and Kraken (NGC\,362, 1261, 3201,
5139, 5272, 5897, 5904, 5946, 6121, 6284, 6544, 6584, 6752, 6864, 6934, 6981, 7006, 7089).
These GCs also have in general large eccentricities and are moving in either prograde or
retrograde orbits (see top-right panel of Fig.~\ref{fig4}). 
Gaia-Sausage is the same  Gaia-Enceladus
elongated structure in velocity space mentioned above, created by a massive dwarf galaxy 
($\sim$ 5$\times$10$^{10}$ $M_\sun$) on a strongly radial  orbit that merged with the MW at 
a redshift $z \la$ 3 \citep{belokurovetal2018}. \citet{myeongetal2018} listed NGC\,362, 
1261, 1851, 1904, 2298, 2808, 5286, 6779,  6864 and 7089 as probable candidate GCs associated to Gaia-Sausage, showing a partial overlap with those listed by \citet{helmietal2018}. We depicted them in the bottom-left
panel of Fig.~\ref{fig4}, showing that they also split into prograde and retrograde
highly eccentric orbits. All candidate GCs with a dwarf origin have semi-major axes
from $\sim$ 5 up to 25 kpc and  $V_\phi$ velocity components relatively small (see bottom-right panel of
Fig.~\ref{fig4}).

At this point, some unavoidable issues arise: firstly, there is an overlap of GCs associated
to different host dwarf galaxies accreted onto the MW. Indeed, by comparing the list of
GCs associated to Gaia-Sausage, Gaia-Enceladus, Sagittarius, Canis Major and Kraken, it is easy to
identify those GCs included in two or three different lists. \citet{myeongetal2018} used 6D 
information to search structures in action  space of 91 GCs and a characteristic energy which 
separates the {\it in-situ} objects in Gaia-Sausage. Similarly, \citet{helmietal2018} 
constrained the azimuthal angular momentum $L_z$ to be smaller than 250 kpc km/s, in addition 
to distances between 5 and 15 kpc from the Sun, and 40$\degr$ away from the Galactic center to
select GCs associated to Gaia-Enceladus. Finally, \citet{kruijssenetal2018} based the selection 
of GCs associated to Sagittarius, Canis Major and Kraken on the reconstruction of the MW's 
merger tree from its GC age-metallicity distribution, and on the estimation of the number of
mergers as a function of mass ratio and  redshift. As can be inferred from the mentioned
works, the partial agreement found between the outcomes of different selection procedures
points to the need of further refinement.

Secondly, every group of associated GCs does not contain only GCs in retrograde orbital motions or in prograde ones, with the exception of Sagittarius. Gaia-Sausage and
Canis Major have the same number of GCs with prograde/retrograde orbits, Gaia-
Enceladus have a prograde/retrograde orbits ratio of 8:5, while Kraken 5:13.
This means that either the selection of GCs associated to accreted dwarf
galaxies based only on their angular momentum, or on their energies or on age-metallicity relationships is not enough as selection criteria. These astrophysical properties
in addition to other properties would seem to be needed. Note, particularly, that
two methods of selecting GCs associated to Gaia-Enceladus (= Gaia-Sausage) have
obtained two different GC samples, with some overlap \citep{helmietal2018,myeongetal2018}.
If we assumed that any applied methods to find out GCs associated to
accreted dwarf galaxies were robust, we should admit that  GCs associated to the
same accreted dwarf could have been deposited in retrograde/prograde orbits randomly.

Thirdly, according to \citet{kruijssenetal2018} the ratio of accreted to {\it in-situ} GCs is 
$\sim$ 2/3, i.e., nearly 40 per cent of the GC population was formed in  dwarf galaxies. 
Here we assumed that the total number of accreted GCs is twice as big as that of GCs
with retrograde orbits -- we assigned the same probability to accreted GCs with prograde/retrograde orbits --, so that the accreted to {\it in-situ} GCs ratio turns out
to be $\sim$ 1. This ratio is $\sim$ 1.5  times that of \citet{kruijssenetal2018}.
Note that the present analysis does not favor GCs mainly being 
formed {\it in-situ}, nor accreted ones being observed only in the outer halo.
It still remains an open issue whether the accreted GC population 
has been shaped by minor mergers \citep[ratio 1:100][]{kruijssenetal2018} or by one major merger 
event \citep[ratio 1:4][]{belokurovetal2018,helmietal2018}.

Recently, several works have pointed out fairly large velocity anisotropy values ($\beta$) 
for the outer halo, and hence have characterized the motion of the halo stellar component
like a more radial than a tangential subsystem. \citet{birdetal2018} obtained $\beta$ $\approx$ 
0.9 over the Galactocentric distance ($R_{GC}$) range 5 - 25 kpc for stars more metal-rich than
[Fe/H] = -1.8 dex, and 0.6 for those more metal-poor
(see also, \citet{cunninghametal2018} ($\beta=0.6$)).
From $R_{GC}$ = 25 kpc up to 100 kpc, \citet{birdetal2018} found that $\beta$ steadily 
decreases until $\sim$ 0.3, independently of the metal content. \citet{belokurovetal2018} agree
with a high $\beta$ value (0.9) for stars more metal-rich than [Fe/H] = -1.7 dex distributed 
within 10 kpc from the Sun. However, they derived a smaller one  (0.2$<$ $\beta$  $<$0.4)
for more metal-poor stars. Summing up, there seems to be a general agreement about 
the value of $\beta$ as a function of the Galactocentric distance and its dependence on 
metallicity. 

As for GCs, \citet{bw2017} found from a modeled GC system  $\beta$ $\sim$
0.68 at $R_{GC}$ = 12 kpc with a steady decrease down to 0.53 at $R_{GC}$ =30 kpc.
\citet{watkinsetal2019} used 34 halo ones with distances to the MW center
between 2.0 and 21.2 kpc and derived $\beta=0.5$.
\citet{vasiliev2018} derived a nearly constant
$\beta$ $\sim$ 0.6 for $R_{GC}$ $>$ 25 kpc, and a decrease in the inwards direction
down to $\beta$ $\sim$ 0.4 at $R_{GC}$ = 5 kpc, and 0.0 at the MW center. The constant
trend outwards 25 kpc does not match the decrease found by \citet{birdetal2018}, while
for $R_{GC}$ $<$ 25 kpc, his value resembles those obtained from field stars more metal-poor
than [Fe/H] $<$ -1.7 dex \citep{birdetal2018,cunninghametal2018,watkinsetal2019}.
Despite the small difference between the present GCs sample and that used by \citet{vasiliev2018}, 
and the fact that we distinguished between prograde and
retrograde orbits, our $\beta$ values for GCs with prograde orbits are in fairly good
agreement with his. 

\begin{figure}
     \includegraphics[width=\columnwidth]{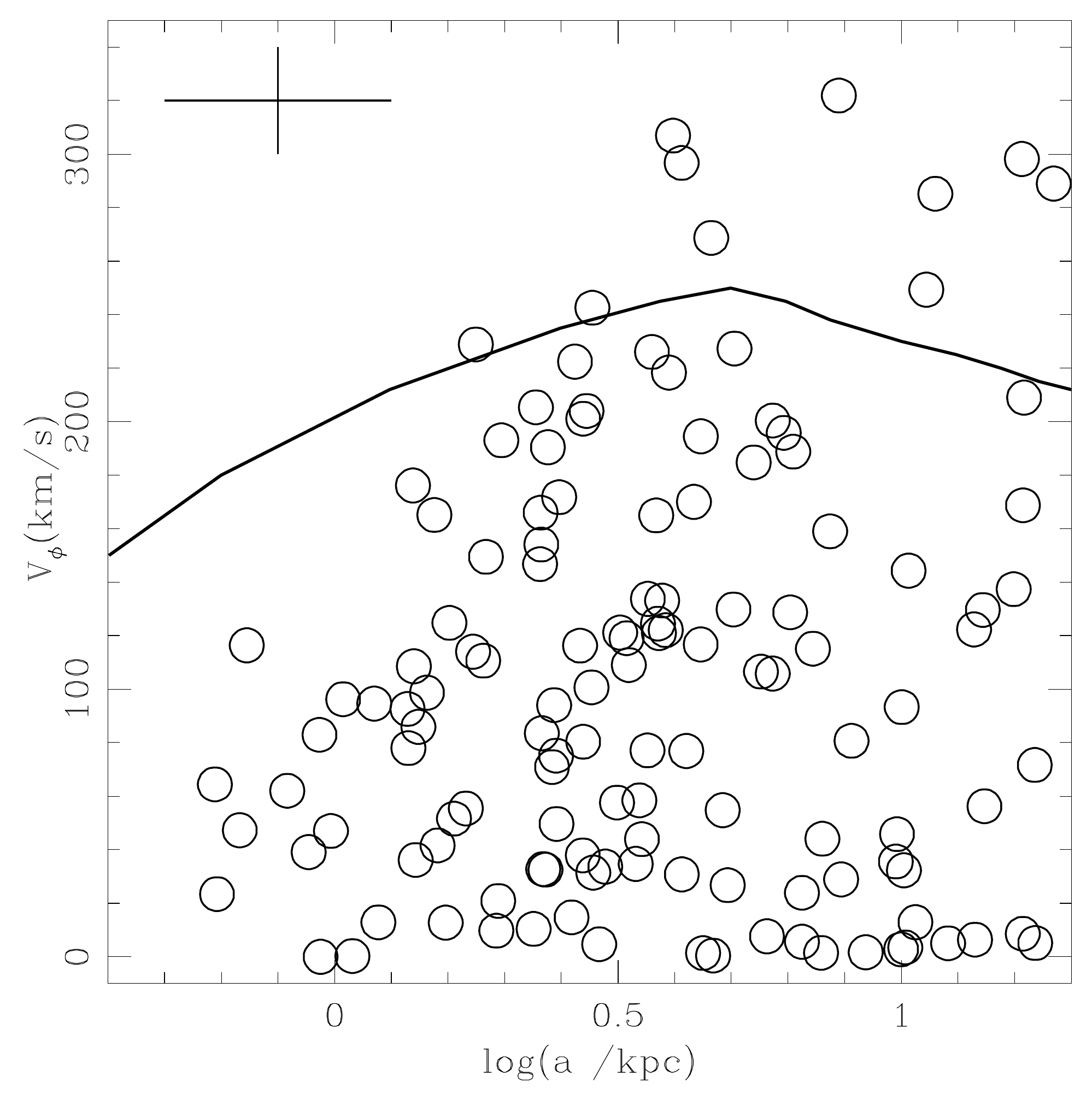}
\caption{ Total circular velocity curve of the MW derived by \citet{crostaetal2018}. Open circles represent GCs in prograde orbits. Typical error bars are also
indicated.}
 \label{fig5}
\end{figure}

Finally, we analyzed the kinematics of the GC population in light of the MW rotation 
curve recently  derived by \citet[][see also figure 16 in Bland-Hawthorn \& Gerhard, 2016]{crostaetal2018} (see Fig.~\ref{fig5}). In the figure, we
considered only GCs in prograde orbits. As can be seen, bulge GCs (log($a$) $<$ 0.2)
do  have velocity components in the direction of the disk rotation smaller that those
predicted for the MW bulge (red line). Here we speculate with the
possibility that  GCs and the MW bulge do not share similar kinematics in the direction of the 
disk rotation or that there still are 
bulge GCs not found \cite[see, e.g.][]{ruyetal2018,camargo2018}. 
For GCs spread throughout the MW's disk, their  velocity components in the direction
of the disk rotation span the whole
range below the total MW rotation curve.  As we mentioned above, they have been formed
from gas that fell increasingly circularized into the growing disk, hence the dispersion in
their circular velocities. There is also a group of GCs that have
$V_\phi$ values higher than $\sim$ 250 km/s. They have eccentric orbits
and fall outside the mean correlation of Fig.~\ref{fig1}, and could have an accretion origin.

\section{Conclusions}

In this work we take advantage of the catalogue of space velocities for nearly all
confirmed MW GC population built recently by \citet{baumgardtetal2019},
aiming at studying the global kinematic properties of them as one of the ancient Galactic 
subsystems. Previous attempts have been constrained by the smaller number of GCs with
accurate proper motions and line-of-sight velocities, among others. 

We show that the relationship between the eccentricity and the inclination of the
GC orbits sheds light on the possible sequence of events that occurred when the
Galactic GC system formed. Although the calculated inclinations refer to the present-day
values, it seems that, for prograte orbits 
 and 3 kpc $<$ $a$ $\le$ 20 kpc, they vary at most $\sim$ 10$\degr$ around 
the mean value at a fixed eccentricity. This behavior makes the inclination of the GC orbit
a useful time-independent orbital parameter. 
Indeed, the resulting linear relationship between the eccentricity and the inclination 
of prograde orbits suggests that the larger the eccentricity the higher the inclination.
This trend resulted to be also a function of the semi-major axis (or averaged Galactocentric 
distance), so that the outermost GCs have the orbits with the highest inclinations respect 
to the Galactic plane and large eccentricity. For GCs with retrograde orbits,
which represent 1/3 of those with prograde orbits, there is
mostly dispersion over the whole inclination range for eccentricities larger than $\sim$ 0.5.

The eccentricity versus inclination relationship for GCs rotating in prograde orbits 
reveals that the initial collapse of the gas that gave birth to the MW GCs was geometrically
radial with preference for relative high angles respect to the Galactic plane. As the gas
reached the growing rotating disk, it became more circular and parallel to it.
Hence, GCs that have been formed in the outskirts of the MW have very eccentric and
highly inclined orbits, whereas those belonging to the disk show direction of movements more
similar to that of the disk. As for GCs with retrograde highly eccentric orbits, they have likely 
an origin of accretion. Nevertheless, we also identified GCs with prograde highly 
eccentric orbits that could have been accreted.

The  more eccentric orbits of the outermost GCs in comparison with those
of the disk GCs, is also supported by the resulting velocity anisotropy ($\beta$). Particularly, 
we computed $\beta$ for three semi-major axis ($a$) ranges, namely: the innermost GCs ($a$ $\le$
3 kpc), GCs spanning the extension of the Galactic disk (3kpc$<$ $a$ $\le$20 kpc) and outer halo GCs
($a$$>$ 20 kpc). We found that $\beta$ decreases from 0.79 down to 0.29 towards the Galactic center for
prograde GCs. In the case of GCs on retrograde orbits, $\beta$ remains nearly constant (0.75).

\acknowledgments

I thank Holger Baumgardt for providing me with his globular cluster database and contributed to
improve the paper   and  the referee for the thorough reading of the manuscript and
timely suggestions to improve it.


\end{document}